\documentclass[pra,prl,twocolumn,superscriptaddress,floatfix,showpacs,10pt]{revtex4}
\usepackage{graphicx}
\usepackage{amssymb}
\usepackage{verbatim}

\begin{document}

\title{A multi-party correlation measure based on cumulant}

\author{D.L. Zhou}
\affiliation{School of Physics, Georgia Institute of Technology,
Atlanta, Georgia 30332, USA}

\author{B. Zeng}
\affiliation{Department of Physics, Massachusetts Institute of
Technology, MA 02139, USA}

\author{Z. Xu}
\affiliation{Center for Advanced Study, Tsinghua University, Beijing
100084, China}

\author{L. You}
\affiliation{School of Physics, Georgia Institute of Technology,
Atlanta, Georgia 30332, USA} \affiliation{Center for Advanced Study,
Tsinghua University, Beijing 100084, China}

\date{\today}
\begin{abstract}
We propose a genuine multi-party correlation measure for a
multi-party quantum system as the trace norm of the cumulant of the state.
The legitimacy of our multi-party correlation measure is explicitly
demonstrated by proving it satisfies the five basic
conditions required for a correlation measure. As an application we
construct an efficient algorithm for the calculation of our measures
for all stabilizer states.
\end{abstract}

\maketitle

\section{Introduction}

Although a composite system can contain several parties, its physical
properties do not necessarily equal to the sum of its parts.
Correlations among different parts in
a composite system are usually invoked to describe the difference
between the physical properties of a composite system and
the sum of its parts. Roughly speaking, the physical properties
for a composite system are reflected both in the sum of the
physical properties of its parts and the correlations (among different parts)
of the composite system. What makes a composite system more interesting
in some sense can be largely attributed to the existence of correlations
among the constituents of the system.

The rapid development of quantum information science in recent
years has called for serious efforts to characterize correlations in a
composite quantum system. Depending on whether quantum nonlocal
resources are needed in the preparation of a quantum state, the
correlations among different parties in a composite system
can be further classified into classical and quantum correlation \cite{Wer89}.
In the language of quantum information science,
quantum correlation is often called quantum entanglement to
emphasize the inseparability of the quantum state of a composite
system into those of its parties.
 Quantum entanglement is widely believed to be
a useful resource in implementing quantum computation
and information tasks \cite{NC00}.

The two-party correlation, especially the two-party quantum
entanglement, has been extensively studied and by now is in many
sense well understood \cite{BDSW96,VP98,HV01,Ved02}. However, very
little is known about quantum entanglement properties for
multi-party systems despite consorted efforts over the last decade.
As we will show through our work below, not only computational but
also conceptual difficulties arise when one attempts to characterize
genuine multi-party correlations \cite{Ved02}. For example, the
two-party correlation of a quantum state $\rho^{(12)}$ is described
by the mutual entropy of the composite state, i.e.,
$S(1:2)=S(\rho^{(1)})+S(\rho^{(2)})-S(\rho^{(12)})$, where $S(\rho)$
is the von Neaumann entropy of the quantum state $\rho$, which is
defined as $S(\rho)=-\mathrm{Tr}(\rho \log_2 \rho)$. The three-party
mutual entropy for a three-state $\rho^{(123)}$, defined accordingly
as
\begin{eqnarray}
S(1:2:3) &=& S(\rho^{(123)})-S(\rho^{(12)})-S(\rho^{(23)})-S(\rho^{(13)})\nonumber\\
&&+S(\rho^{(1)})+S(\rho^{(2)})+S(\rho^{(3)}),
\end{eqnarray}
does not faithfully characterize genuine three-party correlation
because it is known to take on negative
values for some specific three-party quantum states. Vedral
suggested an alternative three-party correlation measure using
the relative entropy defined as \cite{Ved02}
\begin{eqnarray}
S(\rho^{(123)}||\rho^{(1)}\rho^{(2)}\rho^{(3)}) &=& S(\rho^{(1)})
+S(\rho^{(2)})+S(\rho^{(3)})\nonumber\\
&&-S(\rho^{(123)}).
\end{eqnarray}
Although semi-positive definite for any three-party quantum state
and easily generalizable to arbitrary multi-party cases, this
relative entropy measure is not a proper correlation measure. For a
product state $\rho^{(123)}=\rho^{(1)}\rho^{(23)}$ exhibiting only
two-party correlation, the above relative entropy correlation
measure $S(\rho^{(123)}||\rho^{(1)}\rho^{(2)}\rho^{(3)})=S(2:3)$
simply measures two-party correlation, rather than the genuine
three-party correlation. It turns out
$S(\rho^{(123)}||\rho^{(1)}\rho^{(2)}\rho^{(3)})$ measures the total
correlation, the sum of both the two-party and three-party
correlations in the three-party state.

The aim of this article is to provide a genuine multi-party
correlation measure for arbitrary quantum states. Our choice for the
genuine multi-party correlation measure to be shown and justified
below is intimately related to the cumulant (or the Ursell
expansion) of a multi-party density matrix \cite{Sim05,Gar04,Pat96,oneB}.
The research on the cumulant of a multi-party state has had a long
history. Cumulants were first introduced by T. N. Thiele in 1899,
who called them half-invariants \cite{Thi99}. The name of cumulant
is given by R. Fisher and J. Wishart in 1931 \cite{FW31}. It was
first introduced by Ursell into classical physics in 1927
\cite{Urs27}, and Kahn and Ulenbeck gave the quantum mechanical
treatment in 1938 \cite{KU38}.

The cumulant of a multi-party density matrix was known to be related
to multi-party correlation, and was even called the
correlation operator \cite{AWS98}. However, it does not provide a
legitimate measure to compare the correlations in different quantum states.

This article is arranged as follows. In Sec. II, we outline three
preliminaries: First, we will introduce the trace distance between
two Hermitian operators of the same trace. Second, the cumulant of
a multi-party density matrix is introduced in an instructive way.
Third, the conditions for a legitimate multi-party correlation
measure are discussed. In Sec. III, we define our multi-party
correlation measure as the trace norm of the cumulant of the
multi-party density matrix. We further prove that it is a legitimate
multi-party correlation measure by explicitly demonstrate that it satisfies the five
basic conditions for a multi-party correlation measure. We then apply
our correlation measure to study three cases: simple two-party states,
three-party states, and more general multi-party stabilizer states
\cite{BR01,Got97}. Finally, we summarize our results with a discussion.

\section{Preliminaries}

This section contains the three preliminaries for a convenient introduction
of our correlation measure. First, we generalize the
trace distance between two quantum states to the trace distance
between two Hermitian operators with the same trace. An important
property of this distance is briefly reviewed, which constitutes the key
element in proving the legitimacy of the proposed correlation measure. Second, we
briefly review the concept of the cumulant for a multi-party density matrix in an
instructive way. Third, we discuss the basic requirements for a
legitimate multi-party correlation measure.

\subsection{Trace distance}

The trace distance between two Hermitian operators $\rho$ and
$\sigma$ with the same trace is defined by
\begin{eqnarray}
D(\rho,\sigma)\equiv \frac {1} {2} \mathrm{Tr}
\left|\rho-\sigma\right|.
\end{eqnarray}
This definition is a simple extension of the trace distance between
two quantum states \cite{NC00}. Note that there is an alternative
useful expression for the trace distance:
\begin{eqnarray}
D(\rho,\sigma)={\mathrm{max}}_{\scriptstyle{P}} \mathrm{Tr}
(P(\rho-\sigma)),
\end{eqnarray}
where the maximum is taken over all projectors $P$.

It is easy to check that it satisfies the three basic requirements
for a distance, i.e., {\flushleft
\begin{enumerate}
\item $D(\rho,\sigma)=0 \Leftrightarrow \rho=\sigma$,
\item $D(\rho,\sigma)=D(\sigma,\rho)$,
\item $D(\rho,\tau)\le D(\rho,\sigma)+D(\sigma,\tau)$.
\end{enumerate}
}

The trace distance has the following important property, which
is listed here as a theorem.

\textbf{Theorem} $1$: (\textit{Trace-preserving quantum operations
are contractive}) Suppose $\mathcal{E}$ is a trace-preserving
quantum operation. Let $\rho$ and $\sigma$ be Hermitian operators
with the same trace. Then
\begin{eqnarray}
D(\mathcal{E}(\rho), \mathcal{E}(\sigma) )\le D(\rho, \sigma).
\end{eqnarray}

This theorem is a simple generalization of the well-known theorem
where $\rho$ and $\sigma$ are density operators \cite{NC00}. It
plays an important role in proving our correlation measure is
legitimate. The proof of this theorem is omitted here because its
proof is almost the same to that of the original theorem, which
can be found in several reference books.

\subsection{Cumulants}

We introduce an instructive way to understand the cumulant for an
$N$-party density matrix $\rho^{(12\dots N)}$. Suppose we know
all the $(N-1)$-party reduced density matrices, we can
construct a pseudo- $N$-party density matrix
$\tilde{\rho}^{(12\cdots N)}$ such that all of its reduced density
matrices are correct. Then the cumulant of $\rho^{(12\dots N)}$ is
defined by
\begin{eqnarray}
C(\rho^{(12\cdots N)})\equiv \rho^{(12\cdots
N)}-\tilde{\rho}^{(12\cdots N)}.\label{cum}
\end{eqnarray}

The pseudo- $N$-party density matrix $\tilde{\rho}^{(12\cdots N)}$
can be directly constructed by the following method. First, we find
all the partitions of $T_N\equiv\{12\cdots N\}$. Any partition can
be denoted by $\{S_1,S_2,\cdots, S_M\}$ where $M$ $(\ge 2)$ is the
number of partition, $\forall i,j\in T_M$, $S_i\neq \emptyset$,
$S_i\cap S_j=\emptyset$, and $\prod_{i=1}^{M}\cup S_i=T_N$. Second,
the pseudo- $N$-party density matrix is expressible in the form
\begin{eqnarray}
\tilde{\rho}^{(12\cdots
N)}=\sum_{\{S_i\}}a_{\{S_i\}}\prod_{i}\rho^{(S_i)},\label{pseudo}
\end{eqnarray}
where $a_{\{S_i\}}$ are constants dependent on the specific
partition $\{S_i\}$. Third, all the constants can be determined by
the conditions $\mathrm{Tr}_i \tilde{\rho}^{(12\cdots
N)}=\mathrm{Tr}_i \rho^{(12\cdots N)}$, or
\begin{eqnarray}
\mathrm{Tr}_i C(\rho^{(12\cdots N)})=0,\label{cond}
\end{eqnarray}
for any $i\in T_N$.

To determine the constants $a_{\{S_i\}}$ in Eq. (\ref{pseudo}), we
will use the cumulants to expand the density matrix, which is the
so-called the Ursell expansion of the density matrix \cite{Pat96}.
Let us denote $\rho^{(1)}=C^{(1)}$ and $C(\rho^{(12\cdots
N)})=C^{(12\cdots N)}$. From Eqs. (\ref{cum}) and (\ref{pseudo}), the
density matrix
\begin{eqnarray}
\rho^{S}=\sum_{\{S_i\}}b_{\{S_i\}}\prod_{i}
C^{(S_i)}+C^{(S)},\label{urs}
\end{eqnarray}
where the sum is taken over all the partitions $\{S_i\}$ of $S$ with
$S\subseteq T_N$. Using the condition (\ref{cond}) and the
mathematical induction method, we can prove that the constants
$b_{\{S_i\}}=1$. Using Eq. (\ref{urs}), we obtain the unique
solution of the constants in Eq. (\ref{pseudo}) satisfying the
condition (\ref{cond}): $a_{\{S_i\}}=(-1)^M (M-1)!$
\cite{Gar04,Pat96,Sim05}.

The cumulant of an $N$-party density matrix has the following
important property.

\textbf{Theorem} $2$: If an $N$-party density matrix is a product
state, i.e., $\rho^{(12\cdots N)}=\rho^{(S_1)}\rho^{(S_2)}$, where
$\{S_1,S_2\}$ is a partition of $T_N$, then $\tilde{\rho}^{(12\cdots
N)}=\rho^{(12\cdots N)}$, i.e., the cumulant $C(\rho^{(12\cdots
N)})=0$.

\textbf{Proof}: $N=2$, $C(\rho^{(12)})$ follows obviously from
$\rho^{(12)}=\rho^{(1)}\rho^{(2)}$. Assume the theorem is valid for
all $k\le N-1$, we need to prove it is also valid for $k=N$.
According to Eq. (\ref{urs}), we obtain
\begin{eqnarray}
\rho^{(12\cdots N)}&=&\sum_{\{S_{1i}\}}\prod_{i}
C^{(S_{1i})}\sum_{\{S_{2j}\}}\prod_{j} C^{(S_{2j})}+C^{(12\cdots N)}\nonumber\\
&=&\rho^{(S_1)}\rho^{(S_2)}+C^{(12\cdots N)},\nonumber
\end{eqnarray}
where $\{S_{1i}\}$ and $\{S_{2i}\}$ are partitions of $S_1$ and
$S_2$ respectively. The above equation implies $C^{(12\cdots N)}=0$.
This completes our proof.

\subsection{Conditions of a genuine multi-party correlation measure}

From a general physical consideration, a genuine $N$-party
correlation measure $M_C(\rho^{(12\cdots N)})$ should satisfy the
following five conditions \cite{HV01}.
\begin{enumerate}
\item Negative correlation has no
physical interpretation. $M_C(\rho^{(12\cdots N)})\ge 0$.
\item Any product state implies no genuine $N$-party correlation. If an
$N$-party density matrix $\rho^{(12\cdots
N)}=\rho^{(S_1)}\rho^{(S_2)}$, where $\{S_1,S_2\}$ is a partition of
$T_N$, then $M_C(\rho^{(12\cdots N)})=0$.
\item The correlation measure
is invariant under local unitary transformations.
$M_C(U_L\rho^{(12\cdots N)}U_L^{\dagger})=M_C(\rho^{(12\cdots N)})$,
where $U_L=\prod_{i=1}^{N}U^{(i)}$.
\item The correlation measure is
invariant when the system is augmented by locally non-correlated auxiliary sub-systems.
$M_C(\rho^{(12\cdots N)}\otimes \sigma_L^{(12\cdots
N)})=M_C(\rho^{(12\cdots N)})$, where $\sigma_L^{(12\cdots
N)}=\prod_{i=1}^{N} \sigma^{(i)}$.
\item The
correlation measure is non-increasing under local operations.
$C_M(\mathcal{E}_L(\rho^{(12\cdots N)}))\le C_M(\rho^{(12\cdots
N)})$, where $\mathcal{E}_L=\prod_{i=1}^{N}\mathcal{E}^{(i)}$.
\end{enumerate}

Here we emphasized that condition $2$ is stronger than the following
condition $N$-product version,
\begin{itemize}
\item[$2^{\prime}$.] If an $N$-party density matrix $\rho^{(12\cdots
N)}=\prod_{i=1}^{N}\rho^{(i)}$, then $M_C(\rho^{(12\cdots N)})=0$.
\end{itemize}
As we have mentioned in the introduction, this condition can be used
to define the total correlation, which includes different types of
correlations in the state.

We include the extra condition $4$ as a general requirement
for a legitimate correlation measure because correlations
in a system should not depend on the rest of the world
or ancillary systems ($\sigma_L^{(12\cdots
N)}$) if they are independent and uncorrelated
($\sigma_L^{(12\cdots N)}=\prod_{i=1}^{N} \sigma^{(i)}$).

An optional requirement for a legitimate correlation measure is
related to the so-called additivity, i.e., requiring $M_C(\rho^{(12\cdots
N)}\otimes \sigma_L^{(12\cdots N)})=M_C(\rho^{(12\cdots
N)})+M_C(\sigma^{(12\cdots N)})$ for an absolute correlation scale.
This additivity requirement is clearly stronger than our proposed condition $4$.
We feel such a strong condition is not needed as argued previously
in the basic requirements of
two-party entanglement measure \cite{VP98}.

\section{A genuine multi-party correlation measure}

In this section, we present our central result of a genuine
multi-party correlation measure. It is first proposed and further proved
to satisfy the aforementioned five basic requirements for a multi-party correlation
measure, thus it constitutes a legitimate multi-party correlation measure.
We end this section by demonstrating the applications of our proposed
measures to several important class of examples.

\subsection{General formalism}

\textbf{Definition}: An $N$-party correlation measure of the state
$\rho^{(12\cdots N)}$ is proposed as
\begin{eqnarray}
M_C(\rho^{(12\cdots N)}) &\equiv& D(\rho^{(12\cdots N)},
\tilde{\rho}^{(12\cdots N)})\nonumber\\
&=&\frac {1} {2} \mathrm{Tr}
\left|C(\rho^{(12\cdots N)})\right|,\label{cormeas}
\end{eqnarray}
which constitutes a legitimate genuine multi-party correlation
measure because of the following theorem, which is the main
result of our paper.

\textbf{Main Theorem}: $M_C(\rho^{(12\cdots N)})$ is a legitimate
$N$-party correlation measure, i.e., it satisfies the five basic
conditions for an $N$-party correlation measure.

\textbf{Proof}: Let us prove the five conditions respectively.
\begin{enumerate}
\item $M_C(\rho^{(12\cdots N)})=\frac {1} {2} \mathrm{Tr}
\left|C(\rho^{(12\cdots N)})\right|\ge 0$.
\item $\rho^{(12\cdots N)}=\rho^{(S_1)}\rho^{(S_2)}$ $\Rightarrow$
$C(\rho^{(12\cdots N)})=0$ $\Rightarrow$ $M_C(\rho^{(12\cdots
N)})=0$.\\ Note that we have used Theorem $2$ in the first step of
this proof.
\item Under the action of $U_L$,
\[
\rho^{(S_i)}\mapsto \mathrm{Tr}_{T_N-S_i} (U_L \rho^{(12\cdots N)}
U_L^{\dagger})=U_{S_i} \rho^{(S_i)} U_{S_i}^{\dagger},
\]
where $U_{S_i}=\prod_{j\in S_i} U^{j}$. Using the expression
(\ref{pseudo}) for the psedo-density matrix, we obtain
\[
\tilde{\rho}^{(12\cdots N)}\mapsto U_L \tilde{\rho}^{(12\cdots N)}
U_L^{\dagger}.
\]
Thus, the cumulant
\[
C(U_L\rho^{(12\cdots N)}U_L^{\dagger})=U_LC(\rho^{(12\cdots
N)})U_L^{\dagger}.
\]
Therefore,
\begin{eqnarray}
M_C(U_L\rho^{(12\cdots N)}U_L^{\dagger}) &=& \frac {1} {2} \mathrm{Tr}
\left|U_LC(\rho^{(12\cdots
N)})U_L^{\dagger}\right| \nonumber\\
&=&M_C(\rho^{(12\cdots N)}).
\end{eqnarray}
\item It is easy to prove that
\[
C(\rho^{(12\cdots N)}\otimes \sigma_L^{(12\cdots
N)})=C(\rho^{(12\cdots N)})\otimes \sigma_L^{(12\cdots N)}.
\]
Thus
\begin{eqnarray}
M_C(\rho^{(12\cdots N)}\otimes \sigma_L^{(12\cdots N)}) &=& \frac {1}
{2} \mathrm{Tr} \left|C(\rho^{(12\cdots N)})\otimes
\sigma_L^{(12\cdots N)}\right| \nonumber\\
&=&M_C(\rho^{(12\cdots N)}).
\end{eqnarray}
\item Under the action of $\mathcal{E}_L$,
\[
\rho^{(S_i)}\mapsto \mathrm{Tr}_{T_N-S_i} (\mathcal{E}_L(
\rho^{(12\cdots N)}) )=\mathcal{E}_{S_i} (\rho^{(S_i)}),
\]
where $\mathcal{E}_{S_i}=\prod_{j\in S_i} \mathcal{E}^{(j)}$. Using
the expression (\ref{pseudo}) for the psedo-density matrix, we
obtain
\[
\tilde{\rho}^{(12\cdots N)}\mapsto \mathcal{E}_L
(\tilde{\rho}^{(12\cdots N)}).
\]
Therefore,
\begin{eqnarray}
M_C(\mathcal{E}_L(\rho^{(12\cdots N)})) &=&
D(\mathcal{E}_L(\rho^{(12\cdots N)}),
\mathcal{E}_L(\tilde{\rho}^{(12\cdots N)})) \nonumber\\
&\le &  M_C(\rho^{(12\cdots
N)}).
\end{eqnarray}
Note that Theorem $1$ was used for the proof of the last
inequality.
\end{enumerate}

\subsection{Applications}

\subsubsection{Two-party correlation}

The two-party correlation measure is defined as
\begin{eqnarray}
M_C(\rho^{(12)})=\frac {1} {2} \mathrm{Tr}
|\rho^{(12)}-\rho^{(1)}\rho^{(2)}|.
\end{eqnarray}
The physical meaning of this measure is the distance between the
state $\rho^{(12)}$ and its reduced product state
$\rho^{(1)}\rho^{(2)}$ \cite{NC00}.

Let us apply the two-party correlation measure to the following two
typical states. The first state is the maximally classical
correlated two-qubit state
\begin{eqnarray}
\rho^{(12)}_c=\frac {1} {2} \left(|00\rangle_{(12)} \;
{}_{(12)}\langle 00| +|11\rangle_{(12)} \; _{(12)}\langle
11|\right),
\label{c2s}
\end{eqnarray}
we find the two-party correlation according to our measure is
given by $M_C(\rho^{(12)}_c)=1/2$. The second state
is the maximally entangled Bell state
\begin{eqnarray}
|B\rangle_{(12)}=\frac {1} {\sqrt{2}} \left(|00\rangle_{(12)}
+|11\rangle_{(12)}\right),
\label{b2s}
\end{eqnarray}
for which our two-party correlation measure gives $M_C(|B\rangle_{(12)}\;
_{(12)}\langle B|)=3/4$. The unique difference between the
Bell state (\ref{b2s}) and the maximally classical correlated two-qubit state
(\ref{c2s}) is the very existence of quantum coherence in $|B\rangle_{(12)}$.
Our result shows that quantum coherence will increase
the two-party correlation \cite{GPW04}.

\subsubsection{Three-party correlation}

When dealing with three-party systems, our correlation measure becomes
\begin{eqnarray}
M_C(\rho^{(123)})=\frac {1} {2} \mathrm{Tr} |C(\rho^{(123)})|.
\end{eqnarray}

We again study several types of typical
three-qubit states. Analogously, the first state is the maximally classical
correlated three-qubit state
\begin{eqnarray}
\rho^{(123)}_c=\frac {1} {2} \left(|000\rangle_{(123)} \;
{}_{(123)}\langle 000| +|111\rangle_{(123)} \; _{(123)}\langle
111|\right), \nonumber\\
\end{eqnarray}
for which our three-party correlation measure vanishes, i.e.,
$M_{C}(\rho^{(123)}_c)=0$.

The second state we consider is the GHZ state \cite{GHZ89}
\begin{eqnarray}
|G\rangle_{(123)}=\frac {1} {\sqrt{2}} \left(|000\rangle_{(123)}
+|111\rangle_{(123)}\right),
\label{eq19}
\end{eqnarray}
for which our three-party correlation measure gives
$M_{C}(|G\rangle_{(123)}\;{}_{(123)}\langle G|)=1/2$.

These two specific examples indicate that just like the case of
two-parties, quantum coherence generally increases the three-party
correlation. More specifically, we note that
$M_{C}(\rho^{(123)}_c)=0$, i.e., there exists no genuine three-party
correlation in this state. Yet, even according to our definition of
separate states included in the condition $2$, this state is NOT
really a non-correlated state. This example thus shows that even for
a three-qubit state, the first of the five conditions becomes a
sufficient but not necessary condition, i.e., if the correlation
measure $M_C(\rho^{(123)})$ is zero, we do not know for sure if this
state is non-correlated or not. Contrary to this for a mixed state,
however, we find that for a three-qubit pure state, the following
theorem is valid.

\textbf{Theorem} $3$: If $\rho^{(123)}$ is a three-qubit pure state,
and the correlation measure $M_C(\rho^{(123)})=0$, then the state is
non-correlated, i.e., it can be written as a direct product of two
density matrixes of mutually independent parts.

The proof of this theorem is attached as an appendix.

Remarkably, this theorem implies the other typical three-qubit pure state,
the so-called W state \cite{DVC00} possesses genuine three-qubit
correlation.

\subsubsection{Multi-party correlations in the $N$-qubit GHZ state}

\begin{figure}[htb]
\includegraphics[width=3.75in]{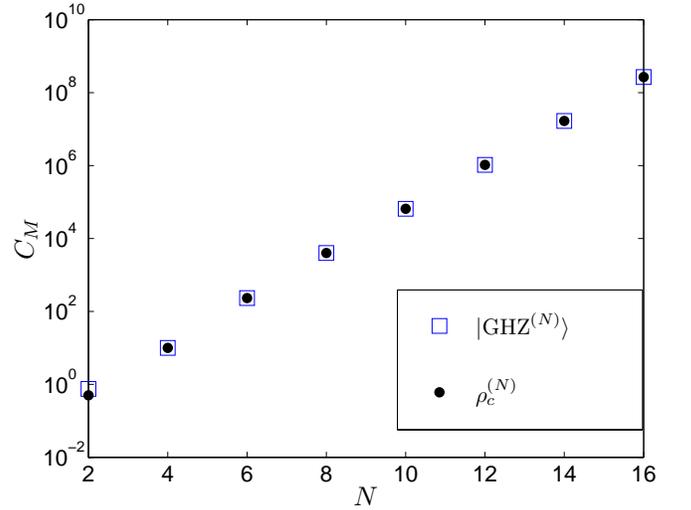}
\caption{(Color online). The $N$-dependence of the total
$N$-partition correlation for an $N$-qubit GHZ state and the maximal
classically correlated $N$-qubit state. }
\label{fig1}
\end{figure}

Let us study the correlations in the $N$-qubit GHZ state, which is
defined by
\begin{equation}
|{\rm GHZ}^{(N)}\rangle=\frac {1} {\sqrt{2}} \left(|0\cdots
0\rangle_{(1\cdots N)} +|1\cdots 1\rangle_{(1\cdots N)}\right).
\end{equation}
An analogous classically correlated state based on what
we considered earlier is
\begin{eqnarray}
\rho_c^{(N)} &=& \frac {1} {2} \left(|0\cdots 0\rangle_{(1\cdots N)}
\; {}_{(1\cdots N)}\langle 0\cdots
0| \right.\nonumber\\
&&+\left.|1\cdots 1\rangle_{(1\cdots N)}\; {}_{(1\cdots N)}\langle
1\cdots 1|\right).
\end{eqnarray}
Since all reduced density matrices of the above two
states are exactly the same, the difference of their respective cumulants
simply consists of the off-diagonal terms
\begin{eqnarray}
&&C(|{\rm GHZ}^{(N)}\rangle)-C(\rho_c^{(N)})\nonumber\\
 &&= \frac {1} {2} \left(|0\cdots
0\rangle_{(1\cdots N)} \; {}_{(1\cdots N)}\langle 1\cdots
1| \right.\nonumber\\
&&\;\;\;+\left.|1\cdots 1\rangle_{(1\cdots N)}\; {}_{(1\cdots
N)}\langle 0\cdots 0|\right).
\end{eqnarray}
Furthermore, the cumulant of the state $\rho_c^{(N)}$ is given by
\begin{eqnarray}
C(\rho_c^{(N)})=c_N\left(\sum\rho_{\textrm{even}}
-\sum\rho_{\textrm{odd}}\right),
\end{eqnarray}
where
\begin{eqnarray}
\rho_{\textrm{even}}&=&\sum_{\{a_i \}\in \{0,1\} }^{\sum_i a_i \in
\textrm{even}}|a_1\cdots a_N\rangle_{(1\cdots N)} \; {}_{(1\cdots
N)}\langle a_1\cdots a_N|,\nonumber\\
\rho_{\textrm{odd}}&=&\sum_{\{a_i \}\in \{0,1\} }^{\sum_i a_i \in
\textrm{odd}}|a_1\cdots a_N\rangle_{(1\cdots N)} \; {}_{(1\cdots
N)}\langle a_1\cdots a_N|,\nonumber
\end{eqnarray}
and the coefficient
\begin{eqnarray}
c_N &=&\sum_{M=1}^{N} \sum_{i=0}^{M-1} \frac {(-1)^{M+i-1} (M-i)^N
(M-1)!} {2^M i! (M-i)!}, \nonumber\\
&=&(-1){\partial^{N-1}\over \partial\lambda^{N-1}}
\left({1\over 1+e^\lambda}\right)\Big|_{\lambda=0}.
\end{eqnarray}
Specifically, for any odd number $N$, $c_N=0$. Therefore for any odd
number $N$, the $N$-party correlation measure gives
\begin{eqnarray}
C_M(\rho_c^{(N)})&=&0,\\
C_M(|{\rm GHZ}^{(N)}\rangle)&=& \frac {1} {2}.
\end{eqnarray}
For an even number $N$, $c_N\neq 0$. For example, we find $c_2=1/4$,
$c_4=-1/8$, $c_6=1/4$, $c_8=-17/16$, $c_{10}=31/4$, $c_{12}=-691/8$,
$c_{14}=5461/4$, $c_{16}=-929569/32$, $\cdots$. The corresponding
$N$-party correlation measure becomes
\begin{eqnarray}
C_M(\rho_c^{(N)})&=&2^{N-1} |c_N|,\\
C_M(|{\rm GHZ}^{(N)}\rangle)&=& 2^{N-1} |c_N|\nonumber\\
&+&\frac {|c_N+\frac {1} {2}|+|c_N-\frac {1} {2}|-2 |c_N|} {2},
\end{eqnarray}
which for $|c_N|\ge 1/2$, gives $C_M(|{\rm
GHZ}^{(N)}\rangle)=C_M(\rho_c^{(N)})$.
In Fig. \ref{fig1}, we show the $N$-dependence of the
above correlation measures for the two states.
Two interesting features are worthy of some attention.
First, it is interesting to note that
$C_M(|{\rm GHZ}^{(N)}\rangle)=C_M(\rho_c^{(N)})$ for
$N\ge 8$ because $|c_N|\ge 1/2$ for
$N\ge 8$. The physical meaning of this interesting equality
is yet to be understood. Secondly, the exponentially
increasing dependence of the total correlations on $N$
in both cases reflects the exponentially increasing size
of the Hilbert space.

\subsubsection{Multi-party correlations in stabilizer states}

The correlations (or entanglement) in stabilizer states have been
discussed in Refs. \cite{HEB04,CYBC04,HIZ05}. We now apply our
multi-party correlation measure (\ref{cormeas}) to characterize
these correlations.

To compute our measure of Eq. (\ref{cormeas}) for a given state, we
need firstly to obtain the corresponding reduced density matrices.
This task is dramatically reduced if the given state is a stabilizer
state. A stabilizer state is the unique simultaneous eigenstate of
the maximal Abelian subgroup of the $N$-qubit Pauli group ${\cal
G}_N$. The group  ${\cal G}_N$ consists of all $4\times 4^N$ local
operators of the form $M= \alpha_M M_1\otimes\cdots\otimes M_N$,
where $\alpha_M\in\{\pm 1, \pm i\}$ is an overall phase factor and
$M_i$ is either the $2\times 2$ identity matrix $I_i$ or one of the
Pauli matrices $X_i$, $Y_i$, $Z_i$.

A stabilizer ${\cal S}$ in the Pauli group ${\cal G}_N$ is defined
as an Abelian subgroup of ${\cal G}_N$ which does not contain $-I$.
A stabilizer consists of $2^k$ Hermitian Pauli operators (i.e. they
must have real overall phase factors $\pm 1 $), where $k \;(\leq N)$
is the number of the generators of $\mathcal{S}$. As the operators
in a stabilizer commute, they can be diagonalized simultaneously
and, what is more, if $k=N$ then there exists a unique state
$|\psi\rangle$ on $N$ qubits such that $M|\psi\rangle =
|\psi\rangle$ for every $M\in{\cal S}$. Such a state $|\psi\rangle$
is called a stabilizer state and the group ${\cal S}={\cal S}(\psi)$
is called the stabilizer of $|\psi\rangle$. The expansion
\begin{eqnarray}
|\psi\rangle\langle\psi| = \frac{1}{2^N}\sum_{M\in {\cal
S}(\psi)}M,\label{Stablizer}
\end{eqnarray}
which describes a stabilizer state as a sum of all elements in its
stabilizer, can be readily verified.

The support supp$(M)$ of an element $M = \alpha_M
M_1\otimes\dots\otimes M_n\in{\cal S}(\psi)$ is the set of all $i\in
\{1, \dots, n \}$ such that $M_i$ differs from the identity $I_i$.
Let $S$ be a subset of $T_N$. Tracing out all qubits of
$|\psi\rangle$ outside $S$ gives the reduced density matrix
$\rho^{(S)}(\psi)$ associate with $S$, which is equal to
\begin{eqnarray}
\rho^{(S)}(\psi)=\frac{1}{2^{|S|}}\sum_{M\in{\cal S}, \mbox{
\scriptsize supp}(M)\subseteq S} M.\label{Reduced}
\end{eqnarray}
This can easily be verified using the identity (\ref{Stablizer}).
This allows use to obtain
the cumulant of the state by direct computation,
with the result taking the following form
\begin{eqnarray}
C(|\psi\rangle \langle \psi|)=\frac {1} {2^N} \sum_{M\in
\mathcal{S}} \beta_M M,
\end{eqnarray}
where $\beta_M$ are real constants, and $\beta_M=0$ when
$M=\prod_{i=1}^{N} I^{i}$. The eigenvalues of the cumulant
$C(|\psi\rangle \langle \psi|)$ can be obtained as follows. First we
take a generator $S_G$ of the stabilizer group $\mathcal{S}$, which
includes $n$ independent elements in $\mathcal{S}$, denoted by $G_i
\;(i=1,2\cdots, n)$. We take values of all the $G_i$ as $\pm 1$,
which determines the values $v_{\{G_i\}}(M)$ of all the $M$. All the
eigenvalues of the cumulant are
\[\frac
{1} {2^N} \sum_{M\in \mathcal{S}} \beta_M v_{\{G_i\}}(M).
\]
Therefore, our multi-party correlation measure for a stabilizer state becomes
\begin{eqnarray}
M_C(|\psi\rangle \langle \psi|) = \frac {1} {2^{N+1}}
\sum_{\{G_i\}}\left| \sum_{M\in \mathcal{S}} \beta_M
v_{\{G_i\}}(M)\right|.
\end{eqnarray}
Without any complication,
the above procedure we describe is easily extended for different
partitions of the stabilizer states, or a reduced density matrix of
the stabilizer states.

Let us demonstrate the above procedure with the three-qubit
stabilizer state $(N=3)$, or the equivalent three-qubit GHZ state
as considered earlier in Eq. (\ref{eq19}). When viewed as
a stabilizer state, its generators can be taken as follows.
\begin{eqnarray}
G_1=X_1 Z_2,\; G_2=Z_1 X_2 Z_3,\; G_3=Z_2 X_3.\nonumber
\end{eqnarray}
The elements in the stabilizer group $\mathcal{S}$ can be written as
\begin{eqnarray}
M_{a_1 a_2 a_3}&=&G_1^{a_1} G_2^{a_2} G_3^{a_3}\nonumber\\
&=& X_1^{a_1} Z_1^{a_2} Z_2^{a_1} X_2^{a_2} Z_2^{a_3} Z_3^{a_2}
X_3^{a_3},\nonumber
\end{eqnarray}
where $a_1, a_2, a_3 \in \{0,1\}$. So the three-party density matrix
is
\[
\rho^{(123)}=\frac {1} {2^3} \sum_{a_1,a_2,a_3=0}^{1} G_1^{a_1}
G_2^{a_2} G_3^{a_3}.
\]
The two-party density matrices are
\begin{eqnarray}
\rho^{(12)}=\frac {1} {2^2} \sum_{a_2=a_3=0} G_1^{a_1} G_2^{a_2}
G_3^{a_3}=\frac {1} {2^2} \sum_{a_1} G_1^{a_1},\nonumber
\end{eqnarray}
\begin{eqnarray}
\rho^{(23)}=\frac {1} {2^2} \sum_{a_1=a_2=0} G_1^{a_1} G_2^{a_2}
G_3^{a_3}=\frac {1} {2^2} \sum_{a_3} G_3^{a_3},\nonumber
\end{eqnarray}
\begin{eqnarray}
\rho^{(13)}=\frac {1} {2^2} \sum_{a_2=a_1+a_3=0} G_1^{a_1} G_2^{a_2}
G_3^{a_3}=\frac {1} {2^2} \sum_{a_1+a_3=0}
G_1^{a_1}G_3^{a_3},\nonumber
\end{eqnarray}
and the one-party density matrices are
\[\rho^{(1)}=\frac {I_1} {2},\;
\rho^{(2)}=\frac {I_2} {2},\; \rho^{(3)}=\frac {I_3} {2}. \]
The cumulant thus becomes
\begin{eqnarray}
C(\rho^{(123)})=\frac {1} {2^3} (G_1 G_2 G_3+G_1G_2 +G_2 G_3
+G_2),\nonumber
\end{eqnarray}
and our three-party correlation measure is given by
\begin{eqnarray}
C(\rho^{(123)})&=&\frac {1} {2^4} \sum_{G_1,G_2,G_3=\pm 1}\nonumber\\
&& |G_1 G_2
G_3+G_1G_2 +G_2 G_3 +G_2|=\frac {1} {2},\nonumber
\end{eqnarray}
the same result as obtained earlier in the line after Eq. (\ref{eq19})
based on a direct calculation.

\section{Discussion and conclusion}

We proposed a multi-party correlation measure as the trace norm of
the cumulant of the multi-party density matrix, as defined in Eq.
(\ref{cormeas}). A natural question arises: Is this a unique
multi-party correlation measure based on the cumulant? For example,
by replacing the trace norm of Eq. (\ref{cormeas}) by a squared norm,
can we obtain an alternative measure
\begin{eqnarray}
M^{\prime}_C(\rho^{(1 2 \cdots N)}) =\text{Tr}\left|C(\rho^{(1 2
\cdots N)})\right|^2,
\label{cormeas2}
\end{eqnarray}
which then can be directly written as a sum of a complete set of squared
correlation functions.
We find that $M^{\prime}_C(\rho^{(1 2 \cdots N)})$
satisfies the condition $1$, $2$, and $3$ of a multi-party
correlation measure. In particular, it is invariant under local unitary
transformations, thus it is appropriate to call
$M^{\prime}_C(\rho^{(1 2 \cdots N)})$ a local unitary invariant
multi-party correlation function (LUI-MCF). We do not know whether
the LUI-MCF satisfies the condition $5$, i.e., whether
$M^{\prime}_C$ is non-increasing under general local operations.
Unfortunately, We find it does not satisfy the additional condition $4$
proposed by us, thus
the LUI-MCF is not a legitimate correlation measure.

As mentioned in the introduction, a total correlation was previously
defined by Vedral to measure the total correlation in a multi-party
quantum state \cite{Ved02}. Our work above also suggests an
alternative total correlation measure as the
 distance between the
quantum state and its reduced completely non-correlated state, i.e.
\begin{eqnarray}
M_{TC}(\rho^{(12\cdots N)})=D(\rho^{(12\cdots N)}, \prod_{i=1}^{N}
\rho^{(i)}).
\end{eqnarray}
It is easy to check that it satisfies the conditions $1$, $2^{\prime}$,
and $3-5$, but does not satisfy the condition $2$, which implies it
is indeed a legitimate total correlation measure.

We expect our correlation measure will find applications not only in
quantum information science but also in
many-body physics. This expectation is based on the observation that
the usual correlation functions cannot characterize general correlations
in a multi-party quantum state. On one hand, our
correlation measure recovers to the usual correlation function when
there is only a single nonzero correlation function. On the other
hand, our correlation measure satisfies the basic general
requirements for a legitimate correlation measure, which implies
they will faithfully characterize the multi-party correlation in a quantum
state. To our knowledge, this is for the first time a
multi-party correlation measure capable of capturing genuine
multi-party correlation of a multi-party quantum state is defined.

From a theoretical viewpoint, one open problem is how to extract
the quantum part of the correlation from our correlation measure.
This will then provide a measure of multi-party entanglement.
In addition, as an open question for further investigation,
it will be interesting to find out how our correlation
measure is related to
quantum entanglement measures \cite{CYBC04,HIZ05,oneA} of interest
in quantum information science. In
many-body physics, we are especially interested in finding out
what is really responsible for the quantum phase transition:
Is it the quantum correlation, classical correlation, or the
total correlation \cite{Ost02,WSL04,VPC04,Vid00}.

In summary, we have proposed a multi-party correlation measure based
on the cumulant of multi-party density matrix, which is capable of
characterizing
genuine multi-party correlation. We proved that our correlation
measure is a legitimate multi-party correlation because it satisfies the five
basic requirements for a multi-party correlation measure.
The fourth requirement is suggested by us based on a physical
and operational considerations of multi-party correlation.
As an application, we find an efficient
algorithm to compute the multi-party correlations for all stabilizer
states.

{\vskip 2mm}

We acknowledge interesting discussions with Prof. C.P. Sun and Mr. B. Sun.
This work is supported by NSF and CNSF.

\appendix*
\section{Appendix: Proof of theorem $3$}

Under local unitary transformations, any three-qubit state
can be most economically expressed as
equivalent to \cite{CACLT00}
\[
\left\vert \psi \right\rangle =a_{0}e^{i\phi }\left\vert
000\right\rangle +b_{1}\left\vert 100\right\rangle +b_{2}\left\vert
010\right\rangle +b_{3}\left\vert 001\right\rangle +a_{1}\left\vert
111\right\rangle ,
\]%
where the parameters satisfy
\begin{eqnarray*}
a_{0},a_{1},b_{1},b_{2},b_{3},\phi  &\in &R, \\
a_{0}^{2}+a_{1}^{2}+b_{1}^{2}+b_{2}^{2}+b_{3}^{2} &=&1.
\end{eqnarray*}%
 Let us denote the
cumulant $C(\rho^{(123)})$ as $\mathcal{C}$, i.e.%
\[
\mathcal{C=}\rho^{(123)}-\rho^{(1)}\rho^{(23)}-\rho^{(2)}\rho
^{(13)}-\rho^{(3)}\rho^{(12)}+2\rho^{(1)}\rho^{(2)}\rho^{(3)}.
\]%
We first consider the element%
\begin{eqnarray*}
\langle 100|\mathcal{C}|010\rangle &=&b_{1}b_{2}\left(
1-2a_{0}^{2}\right) \left( a_{1}^{2}+b_{3}^{2}\right) =0.
\end{eqnarray*}%
According to the symmetry of the state, we obtain%
\begin{eqnarray*}
b_{1}b_{2}\left( 1-2a_{0}^{2}\right) \left(
a_{1}^{2}+b_{3}^{2}\right)  &=&0,
\\
b_{2}b_{3}\left( 1-2a_{0}^{2}\right) \left(
a_{1}^{2}+b_{1}^{2}\right)  &=&0,
\\
b_{1}b_{3}\left( 1-2a_{0}^{2}\right) \left(
a_{1}^{2}+b_{2}^{2}\right)  &=&0.
\end{eqnarray*}%
If%
\[
a_{1}^{2}+b_{3}^{2}=0,
\]%
or%
\[
a_{2}^{2}+b_{3}^{2}=0,
\]%
or%
\[
a_{1}^{2}+b_{2}^{2}=0,
\]%
then it is easy to check that the state is a product state.
Therefore we only to
check two cases%
\[
1-2a_{0}^{2}=0
\]%
or%
\begin{eqnarray*}
b_{1}b_{2} &=&0, \\
b_{2}b_{3} &=&0, \\
b_{1}b_{3} &=&0.
\end{eqnarray*}%
We need to further compute the following matrix element
\begin{eqnarray*}
&&\langle 111|\mathcal{C}|111\rangle  \\
&=&a_{1}^{2}a_{0}^{2}\left( 1-2a_{1}^{2}\right)
+2b_{1}^{2}b_{2}^{2}b_{3}^{2}+2a_{1}^{2}\left(
b_{1}^{2}b_{2}^{2}+b_{2}^{2}b_{3}^{2}+b_{1}^{2}b_{3}^{2}\right)  \\
&=&0.
\end{eqnarray*}%
If%
\[
a_{0}^{2}=\frac{1}{2}
\]%
then%
\[
a_{1}^{2}a_{0}^{2}\left( 1-2a_{1}^{2}\right) \geq 0=0,
\]%
which gives
\[
a_{1}^{2}=\frac{1}{2}
\]%
or%
\[
a_{1}=0.
\]%
When $a_{1}^{2}={1}/{2}$,$a_{1}^{2}={1}/{2}$, we find
\[
b_{1}=b_{2}=b_{3}=0,
\]%
which is the GHZ state, whose cummulant takes the maximum value. For $a_{1}=0$%
, the state is a product state. If
\[
a_{0}^{2}\neq \frac{1}{2},
\]%
then
\[
a_{1}^{2}a_{0}^{2}\left( 1-2a_{1}^{2}\right) =0,
\]%
which leads to
\[
a_{1}a_{0}=0,
\]%
or%
\[
a_{1}^{2}=\frac{1}{2}.
\]%
For the former case the state is a product state. Thus we only need
to prove the theorem for the specific state of the latter case
\[
\left\vert \psi \right\rangle =a_{0}e^{i\phi }\left\vert
000\right\rangle +b_{1}\left\vert 100\right\rangle
+\frac{1}{\sqrt{2}}\left\vert 111\right\rangle .
\]%
We only need to check the element%
\[
\langle 000|\mathcal{C}|111\rangle =\frac{1}{\sqrt{2}}a_{0}e^{i\phi
}-a_{0}e^{i\phi }b_{1}b_{1}\frac{1}{\sqrt{2}}=0,
\]%
or
\[
a_{0}\left( 1-b_{1}^{2}\right) =0,
\]%
which gives $a_{0}=0$, i.e., the state is also a product state.



\begin{references}

\bibitem{Wer89} R.F. Werner, Phys. Rev. A \textbf{40}, 4277 (1989).

\bibitem{NC00} M. A. Nielsen and I. L. Chuang, {\it Quantum
Computation and Quantum Information}, Cambridge University Press
(2000).

\bibitem{BDSW96} C.H. Bennett, D.P. DiVincenzo, J.A. Smolin, and
W.K. Wootters, Phys. Rev. A \textbf{54}, 3824 (1996).

\bibitem{VP98} V. Vedral and M.B. Plenio, Phys. Rev. A \textbf{57},
1619 (1998).

\bibitem{HV01} L. Henderson and V. Vedral, J. Phys. A: Math.
Gen. \textbf{34}, 6899 (2001).

\bibitem{Ved02} V. Vedral, Rev. Mod. Phys. \textbf{74}, 197 (2002).

\bibitem{Sim05} B. Simon, \textit{Functional integration and quantum
physics, 2nd}, AMS Chelsea Publishing (2005).

\bibitem{Gar04} C.W. Gardiner, \textit{Handbook of stochastic methods for physics,
chemistry and the natural sciences, 3nd}, Springer (2004).

\bibitem{Pat96} R.K. Pathria, \textit{Statistical physics, 2nd},
Butterworth-Heinemann (1996).

\bibitem{oneB}P. Carruthers, Phys. Rev. A {\bf 43}, 2632 (1991).


\bibitem{Thi99}T.N. Thiele, \textit{Om iagttagelsesl{\ae}ens
halvinvarianter}, Videnskabernes Selskabs Forhandlinger, 135 (1899).

\bibitem{FW31} R. Fisher and J. Wishart, Proceedings of the London Mathematical Society,
Series 2, \textbf{33}, 195 (1931).

\bibitem{Urs27} H.D. Ursell, Proc. Camb. Phil. Soc. \textbf{23}, 685
(1927).

\bibitem{KU38} B. Kahn and G.E. Uhlenbeck, Physica \textbf{5}, 399
(1938).

\bibitem{AWS98} S. Alavi, G.W. Wei, and R.F. Snider, J. Chem. Phys.
\textbf{108} (2), 706 (1998).

\bibitem{BR01}H. J. Briegel and R. Raussendorf,
Phys. Rev. Lett. \textbf{86}, 910 (2001).

\bibitem{Got97} D. Gottesman, \textit{Stabilizer codes and quantum error
correction}, Ph.D. thesis, (1997).


\bibitem{GPW04} B. Groisman, S. Popescu, and A. Winter,
quant-ph/0410091.

\bibitem{GHZ89}D. M. Greenberger, M. Horne, and A. Zeilinger, \textit{Bell¡¯s
theorem, Quantum Theory, and Conceptions of the Universe}, ed. M.
Kafatos, Kluwer, Dordrecht 69 (1989).

\bibitem{DVC00} W. D{\"u}r, G. Vidal, and J. I. Cirac, Phys. Rev. A 62, 062314
(2000).

\bibitem{CACLT00} A. Acin, A. Andrianov, L. Costa, E. Jane, J.I. Latorre, and R.
Tarrach, Phys. Rev. Lett. 85, 1560 (2000).



DiVincenzo, J. Math. Phys. \textbf{43}, 4286 (2002).















\bibitem{HEB04} M. Hein, J. Eisert, and H.J. Briegel, Phys. Rev.
A \textbf{69}, 062311 (2004).




\bibitem{CYBC04} D. Fattal, T.S. Cubitt, Y. Yamamoto, S. Bravyi, and I. L.
Chuang, quant-ph/0406168.

\bibitem{HIZ05} A. Hamma, R. Ionicioiu, and P. Zanardi, Phys. Rev. A \textbf{72},
 012324 (2005).

\bibitem{oneA}S. S. Bullock and G. K. Brennen,
J. Math. Phys. {\bf 45}, 2447 (2004).

\bibitem{Ost02} A. Osterloch \textit{et al.}, Nature
(London) \textbf{416}, 608 (2002).

\bibitem{WSL04}L.-A. Wu, M. S. Sarandy, and D. A. Lidar, Phys. Rev. Lett.
\textbf{93}, 250404 (2004).

\bibitem{VPC04} F. Verstraete, M. Popp, and J. I. Cirac, Phys. Rev.
Lett. \textbf{92}, 027901 (2004).

\bibitem{Vid00} G. Vidal, J. Mod. Opt. \textbf{47}, 355 (2000).





\end{references}
\end{document}